# Projecting Three-dimensional Protein Structure into a One-dimensional Character Code Utilizing the Automated Protein Structure Analysis Method

Sushilee Ranganathan [1], Dmitry Izotov [1], Elfi Kraka [1], and Dieter Cremer *[1,2]

*[1]Department of Chemistry, University of the Pacific, 3601 Pacific Avenue, Stockton, CA 95211, USA, [2] Department of Physics, University of the Pacific, 3601 Pacific Avenue, Stockton, CA 95211, USA.*

*(E-mail: dcremer@pacific.edu)*

**Abstract**: The protein backbone is described as a smooth curved and twisted line in three-dimensional (3D) space and characterized by its curvature κ and torsion τ both expressed as a function of arc length *s*. It is shown that the function τ(s) is sufficient to analyze the contributions of all amino acids to the conformation of the protein backbone. The characteristic peak and trough patterns of the τ(s) diagrams can be translated into a 16-letter code, which provides a rapid identification of helices, strands, and turns, specifies entry and exit points of secondary structural units, and determines their regularity in terms of distortions, kinks or breaks. Via computer encoding, 3D protein structure is projected into a 1D string of conformational letters, which lead to words (secondary structure), combination of words to phrases (supersecondary structure), and finally to whole sentences (representation of protein chains) without loosing conformational details. The 3D-1D-projection procedure represents an extension of the Automated Protein Structure Analysis (APSA) method. APSA has been applied to describe 155 supersecondary structures from 94 proteins and to compare results with Efimov's classification system of supersecondary structure. The applicability of the latter is demonstrated.

## 1. Introduction

Chemists have invented an elaborate vocabulary to describe three-dimensional (3D) shape of molecules and its change. These terms establish the language of conformational analysis and help chemists to quickly inform each other about the conformation of molecules without reverting to computer generated 3D images or solid ball & stick models of molecules. The usefulness and applicability of the conformational language has its limits. In the case of biomolecules, especially proteins, the manifoldness of possible 3D forms is so huge and the interconversions of the latter so complex that without suitable computer representations of biomolecules an understanding of their conformation and 3D shape is hardly possible.

In this work, we extend the language of conformational analysis to proteins in an elegant way of describing their 3D conformations. For this purpose, we describe the protein backbone as a smooth line in 3D space and characterize the shape of the backbone line by curvature $\kappa$ and torsion $\tau$ as a function of the arc length *s* of the line. Each protein leads to characteristic $\kappa$(s) and $\tau$(s) patterns that reveal the existence of loop, helical, or extended structures in the protein. The residue conformations can be encoded into a 16-letter code, which in turn leads to *words, phrases,* and *sentences* representing the conformation of substructures and eventually the conformation of the whole protein. In this way, the complicated 3D structure of a protein is translated into a simple string of letters and words, without loosing shape details of the protein backbone. Such a projection of 3D protein structure into a 'one-dimensional (1D)' letter code (3D-1D projection) is the prerequisite for the automated analysis of protein structure at the tertiary level, the categorization and classification of protein conformations, the collection of structural subunits in libraries, similarity comparisons of proteins, the understanding of folding, and an important step toward the *ab initio* prediction of protein structure.

The method used for the 3D-1D projection is the Automated Protein Structure Analysis (APSA) that we developed recently. [1,2] APSA has been utilized to automate secondary structure assignment in proteins, to determine the extent of residues with ideal helical (63 %) or extended conformational environments (49 %) in the respective secondary structures of a set of 20 proteins, [1] and to categorize distortions in secondary structures by their $\kappa$(s) and $\tau$(s) patterns using the experimental coordinates of 77 proteins. [2] It was shown in a comparison of 20 different domains taken from various levels of the CATH classification system that proteins with the same topology possess similar κ(s) and τ(s) diagrams. [2] In this work, we will extend the applicability of APSA to supersecondary protein structure after introducing the 3D-1D-projection method of APSA.

The description of supersecondary structure implies the analysis of turns, which are the regions of the polypeptide chain in between helices and β-strands. Turns have a history replete with differing classifications and conflicting results. The term turn is used for 'short, well characterized segments often connecting two consecutive secondary structures' [3]. The term random coil was originally used to denote structural units other than α-helices or β-strands. However, it was later realized that these structural regions are not exactly 'random'; they just have φ and ψ values that are non-repeating. Therefore, the term random coils was later replaced by the terms 'loops' for long regions with as of yet undefined and unclassified structure and 'turns' for the small and well characterized segments with non-regular structure. [3] Approximately 50% of residues in proteins belong to either one of these categories. It is one of the goals of protein structure analysis to expand the extent of the turn's 'empire' within loops and eventually remove the need for the latter term. The results of several investigations show that the line between turns and coils is already blurred. Loops up to 13 amino acids have been analyzed and classified in various ways. [4-7]

The turns, not being regular and repeating at their flanks, highlight the differences in the approaches used by each structure analysis method. As early as 1985, "ambiguity" was thought of as a protein's "intrinsic property" and a list of 'reference definitions' and 20 'working definitions' was given that described turns in various ways. [8] Though early investigations were detailed in the understanding of secondary structures [9,10], there remain still some glaring controversies. [11-21] Collecting all independent turn motifs given in literature and using them in harmony to assess the entire protein structure is impractical. What is needed is a unified approach to describe protein structure completely from primary to tertiary level and uniformly including the description of turns. Some studies [5,22-25] suggest ideas to that effect that might result in a much smaller and tractable number of motifs, however a generally applicable classification system is still missing.

Closely connected with the description of turns is the problem of including (excluding) secondary structure boundaries (caps, entries, exits) into turn regions. For our purpose of attempting a unified structure description, however, these problems are immaterial because we will solve them implicitly when investigating supersecondary structures.

**2. Deriving the Letter Code**

The mathematical basis of APSA for calculating curvature and torsion diagrams κ(s) and τ(s) of proteins has been amply discussed in our previous investigations. [1,2] Ideal conformations of $3_{10}$-, α-, and the π-helices as well as ß-strands were identified by setting up ranges (*windows*) of κ(s) and τ(s) values. [1] Modification and extension of these κ–τ windows in the analysis of 77 proteins led to the description of natural helices and ß-strands in terms of body, termini, entry and exit of the secondary structures. [2] Also, a systematic description of distortions, kinks, and breaks in secondary structure became possible. [2]

In the current investigation, we introduce two important modifications and extensions of the κ–τ analysis of APSA. First, we exclusively focus on the torsion τ because this parameter is sufficient to distinguish between secondary structural units. This was already shown in Ref. [1] and shortly discussed for some examples presented in Ref. [2]. The torsion parameter has the advantage to identify the chirality of any helical or (twisted) ribbon-like structure (positive τ values: right-handed twist; negative τ values: left-handed twist) and is more sensitive to conformational changes than the curvature κ, thus making it easy to keep track of conformational changes along the protein backbone.

Secondly, we replace the continuous τ-diagrams by a (continuous) sequence of τ–*windows* where each window is associated with a residue (given by its $C_\alpha$ position) and represented by a letter. This is possible because each residue of the protein backbone leads to a τ-peak, τ-trough or, in rare cases, a τ-base (a flat segment of the backbone line with τ-values close to zero). Each window includes two sets of information that identify each peak (or trough) in a τ-diagram. This is the τ-value at a $C_\alpha$ point ($\tau_\alpha$) and the maximum (minimum) τ-value $\tau_{max}$ ($\tau_{min}$) in the region up to the next $C_\alpha$ atom, which corresponds to the height (depth) of the τ-peak (trough) located in this region. Especially, when $\tau_{max}$ ($\tau_{min}$) values deviate significantly from $\tau_\alpha$ both sets of information help to quickly differentiate between different windows and, by this, different conformations. In the case of ideal α-helices it is sufficient to just indicate the range of possible τ-values in the region between $C_\alpha$ and the next $C_\alpha$ atom belonging to the residue with the next higher number (Table 1, Figure 1).

Table 1, Figures 1a and 1b

The assignment of the 16 peaks (troughs) was guided by quantitative and qualitative criteria, based on previous analysis of ideal structures [1] and distorted, deviating structures from natural proteins [2]. The numerical ranges are given in Table 1 with their respective peak shapes as appearing in the schematic τ(s) diagram of Figure 1a. It should be noted that not all combinations of $C_\alpha$, $\tau_\alpha$ and $\tau_{max}/\tau_{min}$ values of the 16 cases in Table 1 can be presented in Figure 1a where only the 16 most commonly occurring peaks have been drawn. Six letters are sufficient to describe the variety of τ-peaks (troughs) reflecting either helical, extended, or intermediate conformations (the shape property). Adding the affixes plus (+) or minus (-) indicates whether the residue in question contributes to a right- or left-handed twist of the backbone (the sign property). The prefix underscore (_) is also used to denote the transition from an extended left- into a helical right-handed twist at the entry point (starter) of a helix. These characters identify 14 peaks. Two additional windows are specified for right-handed helices alone and do not have negative counterparts. In this way a total of 16 letters corresponding to 16 windows of τ-values is obtained. Applying the 16-letter code to 94 proteins (for the complete list, see Supporting Information) led in a few cases to unspecified torsion peaks, which were simply tagged by the letter X. We observed that 'X' peaks are associated with structure breaks and missing amino acid information found in some of the PDB files.

Among the helical torsions, the A+ peak signifies the ideal α-helix conformation. A group of four A+ peaks represents a turn of a helix having α-like pitch and diameter. Slight deviations along the body of the helix (responsible for relaxing ideal windows to suite natural helices [2]) are accounted for by the H+ peak, and the $3_{10}$-helix peak is denoted by '3+'. The left handed α-helix peak 'A-', though technically a 'helix peak' (Table 1), is listed among the 'loop peaks' (Table 1, Figure 1a) because the conformation is rare, unstable, and irregular to form a helix in a protein. The termini of helices are different from the body and it was shown in detail [2] how the backbone could enter via a region with either negative or positive $\tau$-value. The starter entering from a negative $\tau$ segment has a higher torsion maximum than the body of the helix that is already in a positive $\tau$-environment. Therefore there is the need for the '_A' peak denoting entry into an α-helix and the '_3' peak entry into a $3_{10}$-helix from an uncoiled segment. The entry into the helix from a segment of positive torsion would occur via a turn not as tight as that of a helix, but still being helical. This helix starter peak is therefore already in the α-helical window and needs no special rule for recognition.

A turn tighter than that of the $3_{10}$-helix would be identified by a 'J+' peak, which was first recognized in a segment that was J shaped. This peak has higher torsion than the 3+ peak (3 residues per turn), but its $\tau_{max}$ value is < 0.75 $Å^{-1}$ and it smoothly leads into the positive β-strand peak (2 residues per peak) denoted 'B+' with $\tau_{max}$ > 0.75 $Å^{-1}$. In Ref. [2] it was shown that β-strands have positive or negative torsion peaks depending on whether the strand is right- or left-handed by local twisting. The negative β-strand peak (denoted 'B-') has $\tau < -0.75$ $Å^{-1}$ and smoothly merges into the 'J-' peak (which is the negative counterpart of the J+ peak) mainly found in loop regions. In addition to the J peaks, the other peaks commonly occurring in loop regions are labeled N, U, and W, where again they could be positive or negative. 'N' refers to a low torsion value (close to 0) of an almost planar curve, 'U' to a U-shaped peak, and 'W' to a peak with a $\tau$-height twice as large as that of the U peak. It might be contested that an (almost) planar curve cannot be differentiated into positive or negative torsion. The N+ and N- peaks (Figure 1a, Table 1), however, refer to the direction from which the curve is being approached and not to the $\tau$ of the peak itself.

In Figure 1b, the 16 peak letters used by APSA are ordered in a two-dimensional phase diagram where one direction gives the shape property of the 3D conformation (either L: looping or e: extended), and the other direction corresponds to the chirality of twisting (sign property: negative sign: twisted left-handed; positive sign: twisted right-handed). The regions with 3D relevance to structure are given by the overlap of the circles, where only the yellow regions are populated by residue conformations found in proteins. The small ellipses on the left denote the regions of the well-defined helical (h) conformations (α, $3_{10}$, π). They are embedded in the larger ellipses, which represent other looping conformations (with varying degrees of rise per amino acid) found in turn regions of proteins. Similarly, the small and large ellipses on the right contain the well-extended (ß-strand) and slightly less extended residue conformations. Figure 1b helps to comprehend the construction and organization of the 16-letter code and to associate torsion peaks with 3D conformations possessing a given shape and handedness.

The letter code was implemented into the APSA computer program. It is automatically generated as a part of the 3D-1D-projection method when calculating torsion and curvature diagrams for a protein. At this point it has to be mentioned that backbone-encoding procedures based on other approaches have been suggested already in the past. For example, the face of an icosahedron that the polypeptide chain would point to when aligned in a specified way was labeled with a letter code [26] and regular structures were encoded with the same letter when pointing to the same icosahedron face. In other works, the backbone conformation was encoded from the Ramachandran plot sectioned into regions that were represented by letters [22] or Greek symbols [27]. Compared to these procedures, the APSA method has the advantage of being fully computerized, easy and generally to apply, non-local, continuous, quantitative, and graphical in the representation of its results.

## 2.1 Interpreting peaks and their patterns

Each peak is related via its shape and sign property to the exact conformation of the backbone. One can distinguish between peaks for strongly looping (abbreviated L) or relatively extended (e) environments. The notations N, U, W, A, H, 3+, _3, and _A belong to the L-category (independent of sign) and the notations J and B belong to the e-category (Table 1, Figure 1). A peak is related to the actual conformation of a given amino acid where the latter is meant in a more global sense. The specific conformation of a residue is given by the orientations of its backbone bonds and the dihedral angles φ and ψ derived from the bond orientations. Since the APSA representation of the backbone is coarse-grained, residue conformations are given with regard to the neighboring residues (the *conformational environment*) rather than in form of an exact local description.

The peaks of helices and β-strands are analyzed following rigid definitions for ideal and distorted structures. [1,2] However, demarcation of the peaks into U and W (positive or negative) is a matter of convenience and could be ignored if a more average description of the protein backbone is required. This gives APSA and its 3D-1D-projection method a great amount of flexibility. APSA values for a given residue reflect the conformational influence of neighboring amino acids and even that from a third amino acid down the backbone if the latter strongly deviates from the current pattern. Hence, APSA assesses the conformation of amino acids based on the *environment* in which they are placed. The peaks in Table 1 do not hold secondary structure relevance without occurring as a pattern. For example, a 3+ peak signifies a $3_{10}$-helix only in the presence of other 3+ peaks. When occurring in a loop region, for example among W+ or U+, it just indicates a segment of strong curving equivalent to one-third of a $3_{10}$-helix. A reference list of some such environments is given in Table 2 with their significance in 3D and occurrence in proteins.

Table 2, Figure 2

**2.2 Interpreting conformational patterns in form of "words"**

A detailed analysis of secondary structures and their distortions was presented in an earlier investigation [2] that involved the automated analysis of 77 proteins based on their κ(s) and τ(s) diagrams. For the purpose of demonstrating the application of the new letter code to the description of protein structure, Figure 2 gives two examples from the previous analysis. [2] In Figure 2a, the first two β-strands and most of the first α-helix of ubiquitin (1UBQ) is given in the letter code where *words* corresponding to regular secondary structural units (boxed letters) are immediately recognized. Most useful is the fact that their boundaries can be accurately identified and the turn between the secondary structures is given in detail.

Figure 2b shows a kinked α-helix from chain E of bovine heart cytochrome C oxidase (1V54). The J+ τ-peak on leucine 89 that is responsible of the kink is evident. One cannot consider the helix as being broken; there is partial stretching at L89 as described by the higher τ of 'J+', which results in the change in the orientation of the helix axis. In the case of a J- peak (negative τ corresponding to a τ-trough), the helix axis would be strongly kinked and unwound at the position of J- with respect to the rest of the body. If the backbone starts winding again into its helical shape, τ will switch back to positive values, however the axis of this second part will be tilted with respect to that of the first.

The rules for automated recognition of secondary structures from a conformation-encoded backbone are given in Table 3 (for application examples, see Figure 2). APSA determines besides a primary encoding based on the $\tau$(s) values also a secondary encoding based on the patterns in the primary code. A helix is divided into starter, body, and exit residues. Helical residues are labeled with the secondary code ‘AAAA’; the starter is marked with an additional ‘<’ before the ‘A’ and the exit is marked with ‘>’ after the last ‘A’. Distorted helical residues are labeled ‘AD’ at the secondary level. A starter residue can be _A, _3, A+, H+, U+ or W+. If the starter is followed by at least 2 body members, it will be labeled with a ‘<A’ and the body with an ‘A’. When the exit is included, the minimum length of the helix is 4 residues (1 full turn).

Table 3

A regular β-strand is not further divided as done in a helix because its primary code is a string of successive ‘B’s or ‘J’s (alone or mixed). As β-strands may have positive or negative $\tau$, their primary code could be made of B+ and J+ ($\in$ e+) or B- and J- ($\in$ e-), respectively. Noteworthy is that the β-strand described by APSA is not directly comparable to the conventional notation of a β-strand because the former, contrary to the latter, is not required to participate in a β-ladder. Loop regions containing extended conformations will also be treated as β-strands because they are geometrically not different from the β-strands found in β-sheets.

Distorted β-strands are labeled in small letters and have a starter (<b), a body (b), and an exit (b>) that must all have the same sign. The body is allowed to have one non-helical breaker (for a positive (negative) β-strand, the breaker can be any label other than e+ (e-)) and up to 2 helical breakers. This rule is designed to permit a kink in the β-strand. Extended conformations are found even among the loop regions of proteins and these can have numerous kinks and twists. [2] To reduce the number of structural units, small deviations are discounted. These are however always stored in the primary code and can be referred to when a more detailed level of description is required. This possibility is only given when distortions are assessed in a quantifiable way as is the case with APSA. A minimum of 4 residues is required (in total) for distorted strands with 1 breaker and 5 residues for those with 2 breakers. The control over the degree of kinking is established through the sign requirement.

### 3. Describing Supersecondary Structure in Form of "Phrases"

A supersecondary structure is a pair of secondary structures (with the turn in between) described in terms of their relative orientations in 3D space. Prior work in literature to define, classify, and analyze the supersecondary structures found in proteins used different approaches that range from detailed atom-by-atom description [22,23] to vector-based turn description methods [13]. Here, we will test whether i) APSA can provide a simple and meaningful description of supersecondary structures and ii) this description complies with what is accepted in the literature. Since secondary structures are represented by strings of letters arranged in *words*, supersecondary structure, as a result of the 3D-1D-projection method, will be given by *phrases* that consist of two *words* connected by letters that represent the turn. Hence, the description of supersecondary structures implies the unique identification of turns in the secondary structural environment. Therefore, this section will present also for the first time a systematic APSA description of turns.

For the purpose of identifying supersecondary structures it is useful to first revert to an established classification system. We found in this connection that the work of Efimov on supersecondary structures [27-35] is at a level of description that is comparable to that of APSA although conformation is not the only criterion used by Efimov. Efimov collected the backbone dihedral angles of turns flanked by secondary structures and labeled them by Greek symbols representing sections of the Ramachandran diagram containing the angles in question. [27] In addition to this geometry-based criterion, the protein segments were further analyzed for the distribution of amino acids by type and physiochemical properties, primarily hydrophobicity. [27] Different types of packing of secondary structures were described with typical patterns of amino acid distributions among the αα-, αβ-, βα- and ββ-classes. Turns and half-turns were denoted as standard structures based on their conformation and length; supersecondary structures were identified by the secondary structure entry, exit, and the standard structures they contained. Table 4 (see also Figure 3) presents a reduced set of the categories described by Efimov [27-35], which is used here for testing the applicability of APSA to supersecondary structure (for details of the testing set used, see Supporting Information).

Table 4, Figure 3

For 155 examples taken from a set of 94 proteins, the resultant APSA encoding was analyzed by peak identity (referring to the exact peak) and property (referring to the sign and shape, Figure 1b). Some examples illustrate the encoding given by APSA. In the region encoded as "α B- B+ α", the first helix ends with a B- exit, and the second helix has a B+ entry. Similarly, a subclass having a code of "β {-} e+ L- α" refers to a β-strand followed by a turn where the first amino acid has a negative τ (could be 'e' or 'L'), the second is a member of 'e+' (B+ or J+), and the third, the helix entry, is a negative looping member. The exit of a ß-strand is normally not given apart from a few exceptions.

Under the αα class, a total of 10 subclasses were analyzed by APSA (Table 4, see also Figure 3), and this covered 13 protein segments under 6 subclasses of *hairpins*, 20 segments under *corner*, 12 under *L-shapes,* and 3 under *V-shapes*. According to Efimov [27,28,32] the *hairpins* have 3 unique conformations, which is confirmed by APSA (Table 4). Thee different types of turns are recognized (*phrases* α{h-}{h-}α, α B-B+B+α, αB-B-B-B-α} where the first links the two helices just via exit and entry without any additional residue, the second is a B+B+ two-residue turn, and the third, a B-B-B- three-residue turn (the helix exit contrary to the entry residue is considered to belong to the helix [2]). Efimov partitions the hairpin subclasses further by introducing a hydrophobic core and viewing the direction of the backbone turning from this core. This leads to the classification terms "right- and left-turned", which must not be confused with the terms right- and left-handed used by APSA. The latter are purely geometrical terms where the former are based on the relative positioning of the protein backbone with regard to the hydrophobic core, which has no equivalent in APSA. In this connection, it has to be mentioned that, in the case of the *L-structures* of Efimov (Table 4), the *right-* and the *left-turned* subclasses are conformationally different as reflected by two different turn codes (Table 4).

Corner and hairpin structures are connected. The *short corner* (Figure 4) is found to have a 'B- B± e- B-' connecting pattern. The 11 examples of the corner subclass (Table 4) with the 'B-B-e-B-' turn conformation contain, according to the APSA code, also examples of the third *hairpin group with the two-residue turn*, suggesting that the demarcation between a *hairpin* and a *corner* is not clear (Table 4, Figure 4). Two successive helices arranged in-plane form a *hairpin,* and when arranged orthogonally (forcing the helices into different planes) become a *corner*. If the *hairpin* is made increasingly non-planar, it will assume an approximate 'X' shape and then eventually adopt a cross(+)-type shape (which is a corner). This example demonstrates the problem involved in the classification of supersecondary structures. In natural proteins, the presence of a continuum of structures makes strict definitions problematic as they can cover only a narrow conformational range.

The αβ-class in our test set includes 3 *hairpin subclasses* with a total 4 + 3 + 3 = 10 examples and 4 *arch* subclasses with 11 + 6 + 6 + 6 = 29 examples (Table 4). The patterns agreed in all cases except for two examples each of the first and fourth arch group (Table 4). In the former case, both regions had a B-B- rather than a B-B+ turn pattern. In the latter case, one example has a looping 3+ (rather than an e+) peak and in the latter case, two have U- and B+ peaks (rather than L+ at the start of the β-strand). Although these differences are small, they are clearly reflected by the τ peaks of APSA. The loop regions of the first 2 classes of *hairpins* are completely positive in τ and are strongly looping in the middle, whereas *arches*, which do not require the strong reversing of the backbone as the *hairpins*, have more extended conformations in their patterns.

The first arch subclass (3 examples) in the βα-class (Table 4) has a 'β B- α' turn pattern, where the β-strands universally ending in a B- peak before entering the helix. The second *arch* subclass (5 examples) possesses a longer turn pattern of the 'β L+ B- B+ {e-} α' type where, apart from the helix entry, all turn residues have extended τ peaks. An exception is found at residue 52 in 2ATC (Supporting Information) for which the helix entry is B+ rather than e-. The third *arch* subclass (12 examples) resembles the second in exit and turn conformation, but the number of 'e' residues being just 1, is found to be oriented differently in 3D. Among its 12 examples, 2 deviations are found for 7AAT (at residue 106: a W- peak not being a member of the L+ group) and 3LDH (at residue 55: a B+ helix entry rather than B- one).

For the ßß-class, two conformationally distinct subclasses of *hairpins*, one *arch* subclass, and 3 subclasses of *corners* (Table 4) with a total of 48 examples are investigated. Of the first hairpin subclass, 8 and 6 examples are analyzed for the right- and left-turned structures (as seen from the hydrophobic core [27]) and 4 for the structure with the βαββ bulge (Table 4). The second subclass of *hairpins* having a *right-turned* loop conformation contains 7 examples in the test set investigated. The APSA obtains, according to the turn codes, two different subclasses (Table 4) but the grouping is different, which again reflects the fact that Efimov used additional viewing criteria. – In the case of the *ββ-arch* subclass, one example (2PCY, 25-41) has a τ peak at residue 33 that does not comply with the 16-letter code, whereas the rest of the segment agrees with the arch subclass pattern. - The three subclasses of *ββ-corners,* though symbolically represented in Figure 4 as corresponding to two orthogonal β-strands, in reality correspond to an arrangement in which the β-strands are coiled around each other to varying extents. Efimov's [33] choice for the first subclass (5 examples tested) resembles a double-stranded coil. The second and third *corner* subclasses with 5 and 3 examples, respectively, investigated in this work have more complex twists than the first subclass, the strongest features of which are the looping and extended residue pattern reflected by the turn patterns (Table 4). Though the turn patterns seem to be identical, most of the e and L shaped residues have slightly different individual conformations.

Although Efimov based his classification of supersecondary structures on additional criteria than just conformation (geometry), APSA confirms this classification on the basis of the letter-codes and the *phrases* obtained. Also, it is confirmed that a classification of turns is implicitly carried out when describing supersecondary structure. There is no need to characterize turns as independent secondary structural units as was frequently done in the literature. [8,18-21] On the contrary, turns (and loops) are best described when part of a supersecondary structure. When verifying Efimov's classification system, we have also revealed some shortcomings of this system (αα- and ßß-hairpins) and have identified 10 out of 155 examples that differed by one or two residues from Efimov's classification system. We note that this kind of accuracy cannot be provided by any other than a strictly or solely geometrically based classification system.

## 4. Conclusions

In this work, APSA [1,2] has been both simplified and extended in the way that a) all conformational features of the protein backbone are described by just its torsion τ and b) all amino acids of the backbone are conformationally described by a 16 letter code corresponding to 16 unique peak (trough) forms of τ(s). Combination of these letters in form of *words* leads to a simple presentation of 3D secondary structure, combinations of words in *phrases* to supersecondary structure, and finally the adding of all *phrases* to a *sentence* gives the complete 3D structure of the protein as a combination of letters, words, and phrases. In this way, 3D conformations are presented in an accurate and complete 1D letter code as a result of the 3D-1D projection method introduced in this work.

The various levels of encoding of protein structure are automatically done by APSA and provides a rapid visualization of regular and non-regular structural units. The 3D-1D-projection concept was applied to investigate Efimov's system of categorizing supersecondary structure. [27-35] A total of 155 examples (out of a total number of 94 proteins) spread over 26 supersecondary classes belonging to the αα, αβ, βα and ββ main groups were analyzed. The APSA 3D-1D encoding procedure verified Efimov's classifications for the majority of all investigated examples identifying differences only for 10 cases. Supersecondary structures belonging to the same class differ primarily because of differences with regard to their turns, i.e. a description and classification of supersecondary structures implies a description of turns. APSA gives the turns in simple letter codes that can easily be distinguished. But APSA reveals also that all of Efimov's examples were not based on an exact conformational match. By bringing in viewing angles based on the position of the hydrophobic core of a protein, all one achieves is confusion because geometrically there is no difference (and hence is artificial) as shown for one of the αα-hairpin subclasses and the αα-corner.

APSA provides the possibility of improving or extending Efimov's system, however we do not pursue this objective further. Instead we will use the experience gained in this work to develop a new APSA-based classification system of supersecondary (and tertiary) structure. [36]

**Supporting Information**: A table with the 94 proteins, their names, and PDB identification numbers is given in the Supporting Information. Also, curvature and torsion diagrams, $\kappa(s)$ and $\tau(s)$, and the backbone line are shown for each protein investigated.

## Acknowledgements

DC and EK thank the University of the Pacific for support. Support by the NSF under grant CHE 071893 is acknowledged

**Figure Captions**

**Figure 1.** (**a**) A schematic representation of the 16 types of τ-peaks and their associated letter code. Only one possibility out of the numerical ranges given in Table 1 is shown for each of the 16 peaks. Note that, though the left handed α-helix peak A- is technically a "helical" peak, it can be considered among the "loop" peaks owing to the lack of regular left-handed helices in proteins. Dots represent $C_\alpha$ atoms. The τ-region to the right of each dot belongs to the $C_\alpha$ atom in question. (**b**) Grouping of the 16 peak letters in a two-dimensional phase diagram spanned by the two properties of peak shape (relating to looping (L) or extended (e) conformations) and peak sign (relating to left- (-) or right-handed (+) twist of the backbone). The yellow areas of the diagram are accessible to natural residue conformations. α-Peaks and ß-peaks represent conformations in ideal α-helices and ideal ß-strands, respectively. See text.

**Figure 2.** (**a**): The regularity of the secondary structures in the τ-encoded backbone: Residues for the first two β strands and the first α helix (M1 - I30) of ubiquitin (1UBQ) are given together with residue numbers, the 16-letter code (primary code), secondary code, and secondary structure name. (**b**) A kinked α-helix from chain E of bovine heart cytochrome C oxidase (1V54). The kink at L89 is represented in the torsion diagram by a J+ peak. Coordinates are taken from the PDB.

**Figure 3**. Schematic representation of Efimov's supersecondary classes and subclasses used in evaluating APSA. The number of conformations described under each subclass is given in parentheses. A cartoon is sketched to represent the overall shape signified by 'hairpins', 'corners', and 'arches' under each class (blue cylinders: helices; green arrows: ß-strands; yellow

connection lines: turns). The ‘1x2’ under L-shaped structures stands for 1 pair (left- and right-handed) of conformationally distinct subclasses

Table 1. Definition of the 16-letter code via the associated torsion windows. [a]

| Peak label | $\tau$ Ranges [Å$^{-1}$] | Conformational equivalent |
|---|---|---|
| **Helix peaks** | | |
| A+ | $0.058 < \tau < 0.23$ | A segment of an ideal $\alpha$-helix |
| H+ | $0 < \tau < 0.3$; $\tau_{max} > 0.06$ | More natural $\alpha$-helix with relaxed $\tau$ windows. |
| 3+ | $0.06 < \tau < 0.56$; $\tau_{max} > 0.3$ | A $3_{10}$-helix segment |
| _A | $-0.18<\tau_{\alpha}<-0.06$; $0.2<\tau_{max}<0.4$ | $\alpha$-helix starter from a left-handed helix entry |
| _3 | $-0.25<\tau_{\alpha}<-0.06$; $0.4<\tau_{max}<0.56$ | $3_{10}$-helix starter as in _A with higher $\tau$(max) |
| **Beta peaks and troughs** | | |
| B+ | $-0.4<\tau_{\alpha}<0.4$; $\tau_{max}>0.75$ | Right-handed $\beta$-strand segment |
| B- | $-0.4<\tau_{\alpha}<0.4$; $\tau_{min}<-0.75$ | Left-handed $\beta$-strand segment |
| **Loop peaks** | | |
| A- | $-0.23 < \tau < -0.058$ | Segment of an ideal left-handed $\alpha$-helix |
| J+ | $-0.4<\tau_{\alpha}<0.4$; $0.4<\tau_{max}<0.75$ | A J- in the right-handed sense: $\tau$ in between B+ and 3+ |
| J- | $-0.4<\tau_{\alpha}<0.4$; $-0.75<\tau_{min}<-0.4$ | Loop region 3 with greater $\tau$ than W-; less torsion than a left-handed $\beta$ segment |
| N+ | $\tau_{\alpha}>0$; $-0.07 < \tau_{max/min} < +0.07$ | A flat peak approached from a right-handed segment |
| N- | $\tau_{\alpha} \leq 0$; $-0.07 < \tau_{max/min} < +0.07$ | As in N+, approached from a left-handed region. |
| U+ | $-0.4<\tau_{\alpha}<0.4$; $0.07 < \tau_{max/min} < 0.2$ | ~A+ segment occurring among left-handed loop segments |
| U- | $-0.4 <\tau_{\alpha}<0.4$; $-0.2<\tau_{max/min}< -0.07$ | Loop region 1 with greater $\tau$ than A- |
| W+ | $-0.4<\tau_{\alpha}<0.4$; $0.2<\tau_{max}<0.4$ | Stronger than a U+ and $\tau$ < 3+ |
| W- | $-0.4<\tau_{\alpha}<0.4$; $-0.4<\tau_{min}<-0.2$ | Loop region 2 with greater $\tau$ than U- |

[a] The continuous torsion function $\tau(s)$ is dissected in a sequence of $\tau$-windows that identify for each residue represented by its $C_{\alpha}$ atom the corresponding conformational letter. Symbol $\tau_{\alpha}$ denotes the torsion value at the $C_{\alpha}$ atom, $\tau_{max}$ and $\tau_{min}$ the maximum and minimum $\tau$-values in the range $C_{\alpha}$ to $C_{\alpha+1}$. See Table 2, for an explanation of the letter code in terms of peak patterns.

**Table 2**. Reference list of conformational environments expressed in terms of τ-patterns and described by the 16-letter code. [a]

| Peak | Peak in a pattern | Significance of pattern | Occurrence in proteins |
|---|---|---|---|
| A+ | … A+ A+ A+ A+… | Ideal (right-handed) α-helix | Body of a regular α-helix |
| A- | … A- A- A- A- … | One turn of left-handed α-helix | Very rare; in turns and loops, at the most, 3 residues long |
| H+ | … A+ A+ H+ A+ A+… | Slight distortions of the ideal right-handed helix | Most natural α-helices |
| 3+ | … 3+ 3+ 3+… | Ideal $3_{10}$-helix | $3_{10}$-helix body; found in loops, caps, and helix distortions |
| | … A+ A+ 3+ A+ A+… | A bent α-helix | In distorted helices or their caps |
| _3 | … _3 3+ 3+… | One turn of a perfect $3_{10}$ helix | _3 is a common helix entry |
| | … _3 A+ A+ A+ A+… | Helix is α, starter is not. | An N terminal helix cap |
| | … _3 3+ 3+ A+ A+… | One turn of a $3_{10}$ helix preceding an α-helix | A narrow terminus |
| _A | … _A A+ A+ A+ A+… | A complete ideal α-helix | Helix with a starter and a body |
| N+ | … A+ A+ N+ A+ A+… | Quarter of an α-helix turn is flat | Induces a kink in the body; forms a broader terminus |
| | … B+ B+ N+ B+ B+… | β-Strand bent by 90˚ in the same plane of the ß-ribbon | Frequent among β-strand distorters |
| W+ | … A+ A+ W+ A+ H+… | α-Helix less bent than a 3+ distortion | Helix distortion among caps |
| B+ | … B+ B+ B+ B+ B+… | An extended region with right-handed twist | In regular β-strands |
| | A+ A+ A+ A+ B+… | A helix exit conformation | The B+ is part of the helix |
| B- | … B- B- B- B- B-… | An extended region with left-handed twist | In regular β-strands |
| | A+ A+ A+ A+ B- … | A common helix exit conformation | The B- is part of the helix |
| J- | … B- J- B- B- B-… | Essentially "B- B- B- B- B- B-" with slightly less torsion | β-Strand |
| | … B- J- U- B- B-… | Extended→helical→extended pattern (all left-handed) | Common in loop regions |
| J+ | … A+ A+ A+ 3+ J+… | Helix smoothly merging into a loop | J+ is in between 3+ (3 residues per turn) peak and B+ (2 residues per turn) peak in terms of τ-window definitions |
| | … B+ B+ B+ J+ W+ … | β-Strand smoothly merging into loop | As above; the extended peaks grow shorter and into the neighboring τ-window towards helical turns |
| N- | … B- B- N- B- B-… | β-Strand bent by 90˚ in the same plane (as in N+). | Occurs in negative ß-strands only as N- is approached from negative τ direction |
| U- | | N, U, W, J all the way to B have adjacent τ-windows for $\tau_{max}/\tau_{min}$ | Merging any two adjacent categories can be done to increase flexibility |
| U+ | | If $C_\alpha$ atoms occur within the A+($C_\alpha$) range, then A+ H+, U+, 3+, J+, B+ have a $\tau_{max}$ continuum | |

**Table 3.** Rules for the encoding of secondary structures using the 16-letter code. [1]

| Sec. structural unit | Recognition rule By primary code: Peak label | Recognition rule By primary code: Pattern | Sec. code | Comment |
|---|---|---|---|---|
| α-helix | | | | |
| Starter | _A, _3, A+, H+,U+, W+ | | <A | |
| Body | A+, H+, 3+, N+, U+, W+ | At least 2 body members | A | Starter + body + exit = minimum 4 residues = 1 full turn of α-helix |
| Distorter | 3+, N+, U+, W+ | | AD | |
| Exit | The first non-helical residue | | A> | |
| β-strands | | | | |
| Regular | B+, J+ | 3 consecutive peaks | <B B B> | Right-handed β-strand |
| | B-, J- | | | Left-handed β-strand |
| Distorted & left-handed | | Minimum length = 4 (or 5 *) residues | | |
| Starter | B-, J- | | <b b b> | |
| Body | B-, J- | 1 breaker allowed in the body | | Includes kinks in β-strands |
| | | * 2 breakers (∈ h/e+) allowed in the body | | |
| Exit | The last B-, J- | | | |
| Breaker | Any peak other than B-, J- | | | |
| Distorted & right-handed | Rules as for left-handed. Minimum length = 4 (or 5) residues for the presence of 1 or 2 helix breakers<br>NOTE: The sign of the starter, body and exit match. The sign of the breakers do not matter in both cases. | | | |

**Table 4:** List of supersecondary structures investigated by Efimov [27-33] and verified by APSA.

| **Supersecondary category name**[1] | **APSA: primary encoding** [2] | **Number of examples** | **Number of deviations** |
|---|---|---|---|
| **αα** [27,28,32] | | | |
| Hairpin (γε) | α {-} {-} α | 3 | 0 |
| Hairpin ($\gamma\alpha_L\beta$) | α B- B+ B+ α | 5 | 0 |
| Hairpin ($\gamma\alpha_L\beta_P\beta_P$) | α B- B- B- B- α | 5 | 0 |
| Corner | α B- B+ {e-}B- α | 9 | 0 |
| | α B- B- {e-} B- α | 11 | 0 |
| L- structure, right turned | α B+ α | 8 | 1 |
| L- structure, left turned | α U- B+/B- α | 4 | 1 |
| V-shaped structure ($\gamma\beta\beta_P$) | α {e+} B- B- α | 3 | 0 |
| **αβ** [27,29] | | | |
| Hairpin (βαγβ) | α B+ {h+} {e+} β | 4 | 0 |
| Hairpin ($\beta\beta\alpha_L\beta$) | α B-/B+ {L+} B+ β | 3 | 0 |
| Hairpin (1-5 gaps[3]) | α {e+} {e-} {e-} {h} 3+/B+ β | 3 | 0 |
| Arch ($\gamma\alpha_L\beta$) | α B- B+ β | 11 | 2 |
| Arch ($\gamma\alpha_L\beta\alpha\beta\beta$) | α {e-} B-/B+ B-/B+ {L+} β | 6 | 0 |
| Arch ($\gamma\alpha_L\beta\alpha\alpha\gamma\beta$) | α B- B-/B+ {e-} {e-} β | 6 | 0 |
| Arch (γβαββ) | α {e+} B-/B+ {L+} β | 6 | 2 |
| **βα** [27,30] | | | |
| Arch (no turn) | β B-[4] α | 3 | 0 |
| Arch ($\beta\beta\alpha_L\beta\beta$) | β- {L+} B- B+ {e-} α | 5 | 1 |
| Arch (βα/γβ) | β {L+} B- α | 12 | 2 |
| **ββ** [27,31,33] | | | |
| Hairpin ($\beta\alpha\gamma\alpha_L\beta$) Right-turned | β {+} B- B+ β | 8 | 0 |
| Hairpin ($\beta\alpha\gamma\alpha_L\beta$) | β {+} {+} β | 17 | 0 |
| Arch ($\beta\beta_P\beta_P\alpha_L\beta$) | β N-/{L+} B+ β | 10 | 1 |
| Corner Group 1[5] | β {L} {e}… β | 5 | 0 |
| Corner Group 2[6] | β {L} {e}..{L} {L} {e}..{L} {e}.. β [7, 8] | 5 | 0 |
| Corner Group 3[9] | β {L} {e}..{L} {L} {e}..{L} {e}.. β [7, 8] | 3 | 0 |

[1] Names and turn conformation (in parentheses) from Refs [27-35]. [2] Variations in the APSA string code are indicated with a '/' between in the labels. If peaks fall within a class (see Figure 1b), the class is given by {}. The properties defining the classes are e: extended, L: looping, -: negative τ peaks, +: positive τ-

peaks. α and β denote helix and strand, respectively. The letter code following the first α denotes the helix exit and belongs to the helix. – [3] Not a conformational term: refers to 1→5 hydrophobicity gap patterns. - [4] This is a helix entry as well as the exit of the β-strand. [5] Structure resembles a double strand coiled coil. [33] - [6] The first strand has ββ corner group 1 conformation; the second strand has β-bulge or a small standard structure. [33] [7] Segments have a series of successive turns. [8] The pattern indicates similar peaks found among the investigated examples. [9] Both strands bent over 90° with a small standard structure in between. [33]

Figure 1a

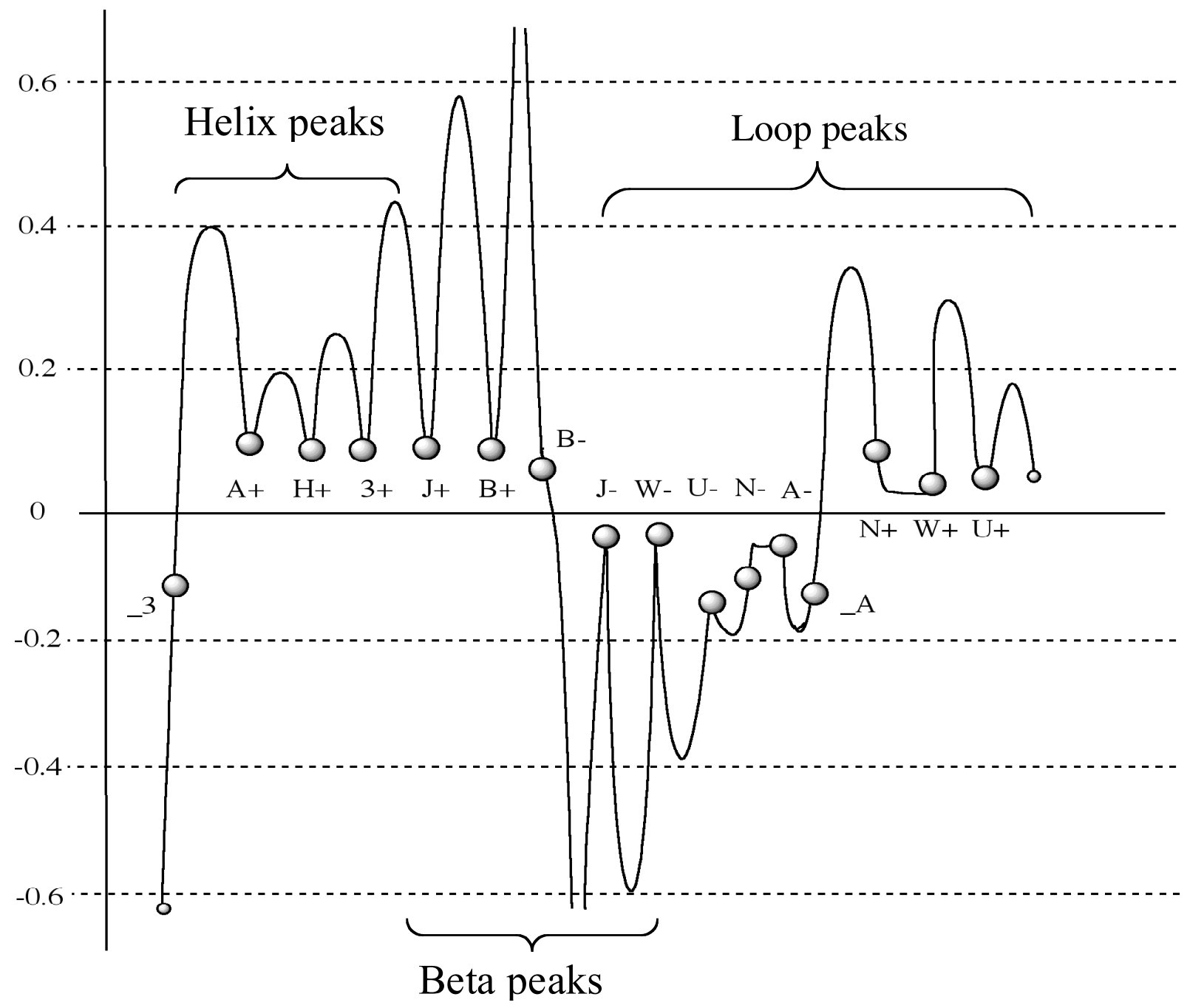

Figure 1b

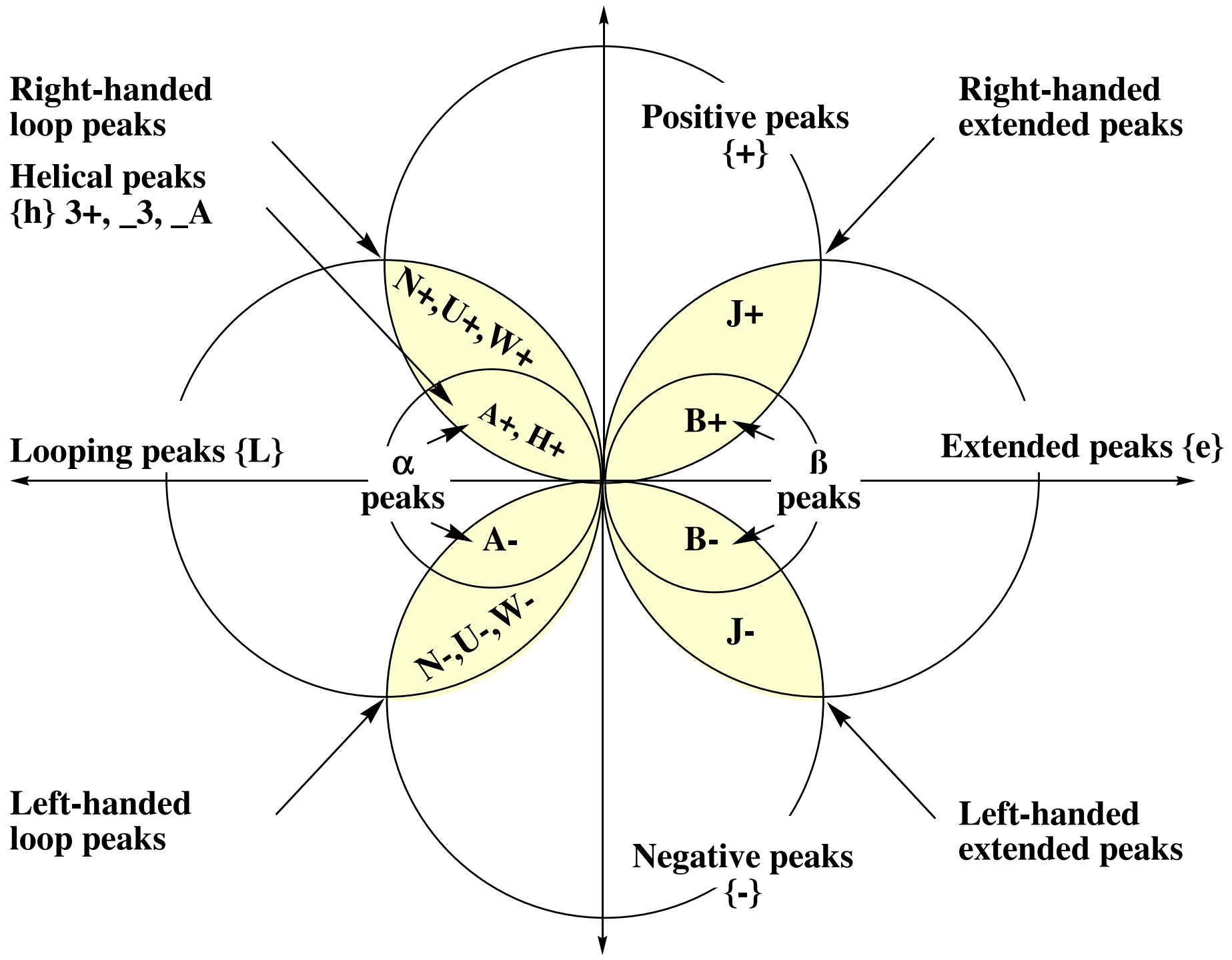

Figure 2a

| | | | | | | | | | | | | | | | |
|---|---|---|---|---|---|---|---|---|---|---|---|---|---|---|---|
| residue name | M | Q | I | F | V | K | T | L | T | G | K | T | I | T | L |
| residue # | 1 | 2 | 3 | 4 | 5 | 6 | 7 | 8 | 9 | 10 | 11 | 12 | 13 | 14 | 15 |
| primary code | B- | B- | B- | B- | B- | B- | B- | J+ | B- | B+ | B- | B- | B- | B- | B- |
| secondary code | <B | B | B | B | B | B | B> | J+ | B- | B+ | <B | B | B | B | B> |
| secondary structure | ß-strand 1 | | | | | | | | | | | ß-strand 2 | | | |
| residue name | E | V | E | P | S | D | T | I | E | N | V | K | A | K | I |
| residue # | 16 | 17 | 18 | 19 | 20 | 21 | 22 | 23 | 24 | 25 | 26 | 27 | 28 | 29 | 30 |
| primary code | B- | B- | B- | _3 | B+ | B- | B- | _3 | A+ | A+ | A+ | A+ | A+ | A+ | A+ |
| secondary code | B | B | B> | _3 | B+ | B- | B- | <A | A | A | A | A | A | A | A |
| secondary structure | ß-strand 2 | | | | | | | | | | | α-helix 1 | | | |

Figure 2b

| | | | | | | | | | | | | | | |
|---|---|---|---|---|---|---|---|---|---|---|---|---|---|---|
| I | Y | P | Y | V | I | Q | E | L | R | P | T | L | N | E |
| 81 | 82 | 83 | 84 | 85 | 86 | 87 | 88 | 89 | 90 | 91 | 92 | 93 | 94 | 95 |
| A+ | A+ | A+ | A+ | A+ | A+ | A+ | A+ | J+ | A+ | A+ | A+ | A+ | A+ | A+ |
| A | A | A | A | A | A | A | A | AD | A | A | A | A | A | A |

primary code

secondary code

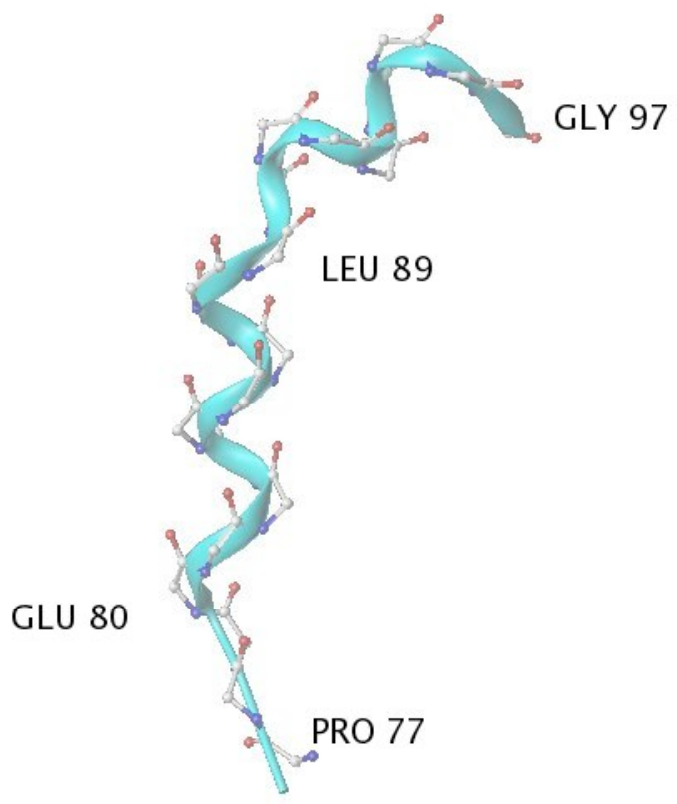

Figure 3

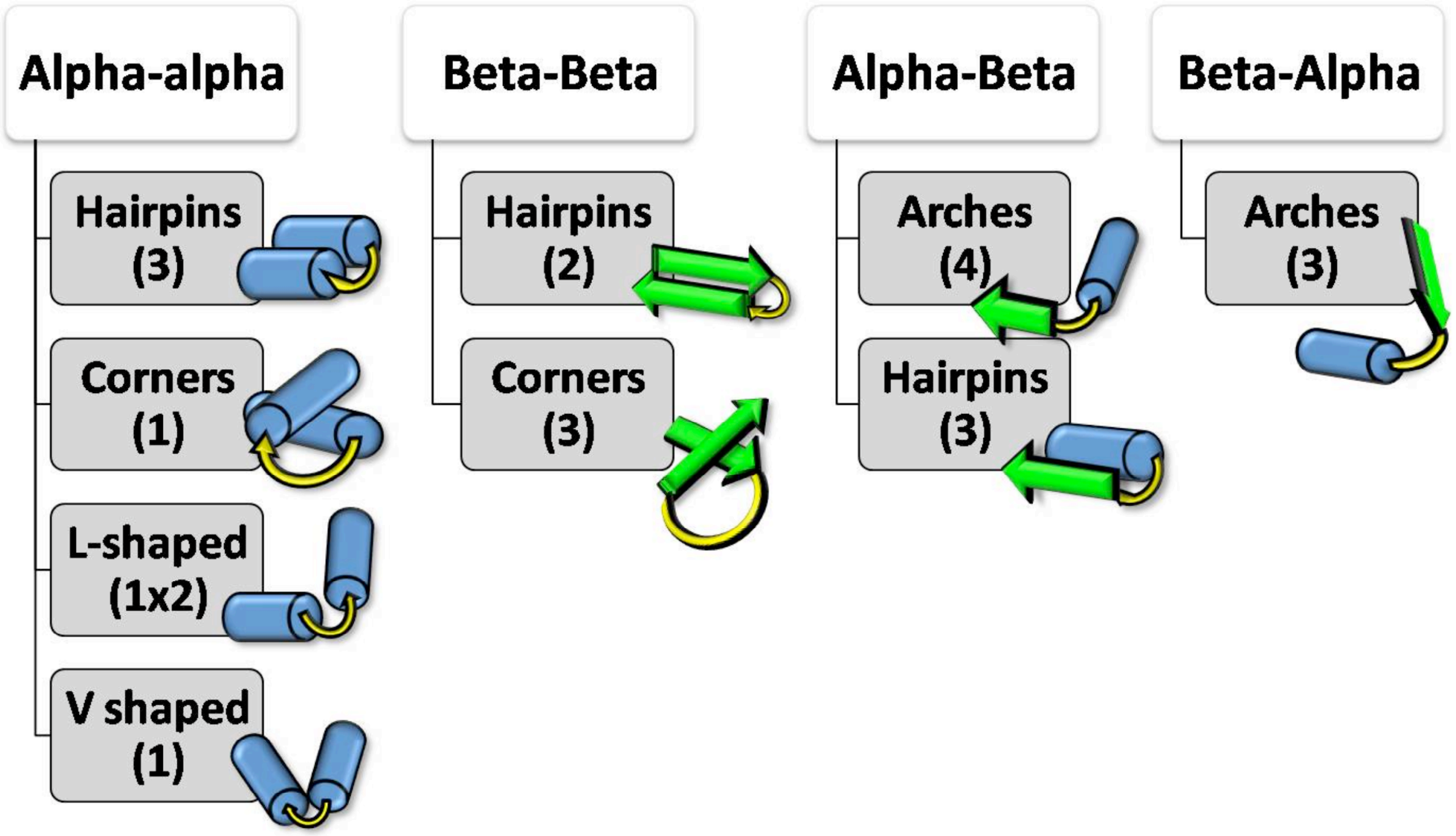